\documentclass[a4paper,fleqn,usenatbib]{mnras}
\usepackage{dcolumn}
\usepackage{bm}
\usepackage{lscape}
\usepackage{graphicx}
\usepackage{subfigure}
\usepackage{hyperref}
\usepackage[hyphenbreaks]{breakurl}
\usepackage{rotating}
\usepackage{tikz}
\usepackage{setspace}
\usepackage{array} 
\usepackage{adjustbox}
\usepackage{siunitx}
\usepackage{float}

\graphicspath{{Pics/}}

\begin{document}

\title[Evolution of the Mass-Size Relation]{The evolution of the stellar mass-size relation of bulges and disks since $z = 1$.}


\author[Hashemizadeh et al.]{Abdolhosein Hashemizadeh$^{1}$\thanks{E-mail: abdolhosein.hashemizadeh@research.uwa.edu.au} 
}

\date{Accepted XXX. Received YYY; in original form ZZZ}

\pubyear{\the\year}


\label{firstpage}
\pagerange{\pageref{firstpage}--\pageref{lastpage}}
\maketitle

\begin{abstract}
We explore the evolution of the stellar mass-size relation of galaxies of different morphological types and specifically bulge and disk components. We use a sample of $\sim35,000$ galaxies within a redshift range $0 < z < 1$, and stellar mass $\log_{10}(\mathrm{M}_*/\mathrm{M}_\odot) \geq 9.5$ volume-limited sample drawn from the combined DEVILS and HST-COSMOS region for which we presented a morphological classification into sub-classes of double-component (BD), pure-disk (pD), elliptical (E), and compact (C) in Paper-I and a structural decomposition into disk (D), diffuse bulge (dB), and compact bulge (cB) in Paper-II.  
We find that compared to disks, ellipticals and bulges follow steeper $M_*-R_e$ relations, likely indicating distinct evolutionary mechanisms. Ellipticals and disk structures follow consistently unchanged slopes of $\sim0.5$ and $\sim0.3$, respectively, at all redshifts. We quantify that disks follow a redshift independent $M_*-R_e$ slope regardless of the presence or absence of a bulge component (i.e., BD or pD) suggesting a similar origin and evolutionary pathway for all disks. Since $z = 1$ compact-bulges present a steepening relation which do not follow that of Es whilst diffuse-bulges experience a modest flattening.
Overall, we find a close-to-no variation in the $M_*-R_e$ relations over the last $\sim8$ Gyr suggesting that despite ongoing although declining star-formation, mass evolution, morphological transitions and mergers, evolution moves galaxies along their $M_*-R_e$ trails. This seems to be consistent with an inside-out growth and evolution picture in which galaxies grow in size as they do in stellar mass. Besides, minor mergers are likely to be responsible for the growth of Es, at least in $z < 1$.  

\end{abstract}

\begin{keywords} galaxies: formation - galaxies: evolution - galaxies: bulges - galaxies: disk - galaxies: elliptical - galaxies: structure - galaxies: general
\end{keywords}

\setlength{\extrarowheight}{0pt}

\section{Introduction}
\label{sec:ch5_intro}

The size and mass evolution of galaxies are arguably two of most important aspects of galaxy formation and evolution, making the scaling relation between these two parameters an informative fundamental plane. The distribution of galaxies on this plane gives insight into their angular momentum and hence their dark matter halos and their assembly history (see e.g., \citealt{Mo97,Mo98}; \citealt{Jiang19}). The mass and angular momentum of disks are argued to be linked to the mass and angular momentum of their halos which are in turn tightly correlated with the density of the Universe at time when the halo was formed (e.g., \citealt{Dalcanton97}). Consequently, this scenario concludes that halos that formed earlier in cosmic time are denser than halos formed at later times that in turn impacts the mass-size relation (e.g., \citealt{Trujillo06}). In addition, galaxies stellar mass-size relations ($M_*-R_e$) changes depending whether built by major or minor mergers (see \citealt{Trujillo07,Trujillo11}). 
Therefore, studying the fundamental plane of mass-size for different galaxy types and components can potentially shed light on how galaxies have reached their current size and mass.  

Disk and spheroid galaxies are shown to follow different $M_*-R_e$ relation (e.g., \citealt{Freeman70}; \citealt{Kormendy79}; \citealt{Kauffmann03}). 
\cite{Shen03} explored the $M_*-R_e$ relation of local galaxies in the Sloan Digital Sky Survey (SDSS, \citealt{York00}) and found that early- and late-type galaxies (ETG and LTG, respectively) follow different relations between their size and stellar mass with LTGs with $M_*/M_\odot \geq 10^{10.6}$ following $R \propto M^{0.4}$ and less massive LTGs following $R \propto M^{0.15}$ while ETGs follow a steeper relation of $R \propto M^{0.56}$. 
Finding a steeper relation for ETGs, they examined two models of (I) single merger event versus (II) multiple mergers for the formation of ETGs and find that the latter scenario can reproduce the observed $M_*-R_e$ relation better than the former so they propose that ETGs are the likely remnants of a sequence of mergers with each merger increasing the size of the galaxy. 
\cite{Lange16} studied the $M_*-R_e$ relation of $z = 0$ galaxies from the Galaxy And Mass Assembly (GAMA, \citealt{Driver11}) survey. They further investigated the relation for galaxies of different morphological types as well as for disks and bulges using structural analysis based on GALFIT (\citealt{Peng02,Peng10}).

The evolution of the $M_*-R_e$ relation has been to interest of both observers and simulators as a key meeting ground between easy observables ($M_*, R_e$) and parameters traced in numerical simulations (e.g., $M_H$). This has become more feasible with the development of deeper photometric and spectroscopic surveys such as Cosmic Assembly Near-infrared Deep Extragalactic Legacy Survey (CANDELS, \citealt{Grogin11}; \citealt{Koekemoer11}) and the Cosmic Evolution Survey (COSMOS, \citealt{Scoville07}). The state-of-the-art study of the evolution of the $M_*-R_e$ with redshift is that performed by \cite{vanderWel14} where they utilized the Hubble Space Telescope (HST) WFC3 imaging of CANDELS and explored the $M_*-R_e$ relations of early- and late-type galaxies between $0 < z < 3$ (also see \citealt{vanderWel12} for their single S\'ersic fit measurements using {\sc GALFIT}, \citealt{Peng02,Peng10}). They highlighted that ETGs evolve faster ($R_{\mathrm{eff}} \propto (1+z)^{-1.48}$) than LTGs ($R_{\mathrm{eff}} \propto (1+z)^{-0.75}$) while the late-type population is larger than the early-type population at all redshifts (\citealt{vanderWel14}). More recently, \cite{Mowla19} combined a sample of high-mass galaxies in the COSMOS-Drift And SHift (COSMOS-DASH) with the 3D-HST/CANDELS sample of \cite{vanderWel14} and provided an updated version of the evolution of the $M_*-R_e$ relation for early- and late-type galaxies.      

The size evolution of disk and spheroid dominated galaxies with redshift is argued to have a direct link to their evolutionary history suggesting that disks that follow a flatter size evolution might have grown through gas infall (e.g., \citealt{Cayon96}; \citealt{Bouwens97}; \citealt{Lilly98}; \citealt{Ravindranath04}; \citealt{Barden05}) while spheroid dominated galaxies following steeper size evolution (e.g., \citealt{vanderWel14} and \citealt{Shen03}) are expected to grow through mostly major mergers (\citealt{White91}). 

As mentioned above, the evolution of the $M_*-R_e$ relation has so far been studied for galaxies in two broad categories (e.g., ETG/LTG). Despite the importance of this fundamental plane the evolution of the $M_*-R_e$ relation of galaxies of different morphological types, and more specifically for bulges and disks components, has not been thoroughly studied in the literature. 
In the present work, we make use of our morphological classifications (as described in \citealt{Hashemizadeh21}; Paper-I) together with our bulge-disk decomposition analysis (as described in \citealt{Hashemizadeh22}; Paper-II), to investigate the evolution of the $M_*-R_e$ relation of galaxies of different morphological types as well as disks and bulges since $z = 1$. 

\section{The evolution of the $M_*-R_{\lowercase{e}}$ relation}
\label{sec:DEVILS_MRe}

In this section, we aim to measure the evolution of the $M_*-R_e$ relation of different morphological types together with that of disks and bulges, separately.
Similar to Paper-I and II, we use our visual morphological classifications and separate our sample into pure-disk, elliptical+compact (E+C, following Paper-II) and double-component (BD) morphological types. We subdivide the BD systems into disk and bulge components using our bulge-disk de-compositions (as described in Paper-II) with bulges then divided into diffuse- and compact-bulges (dB and cB) according to their surface stellar mass densities as described in Section 4 of Paper-II. Note that we use the same dB/cB separation method for both D10/ACS and the $z \sim 0$ GAMA samples. 

\begin{landscape}
\begin{figure} 
	\centering
	\includegraphics[width = 1.3\textwidth, height = 0.97\textwidth]{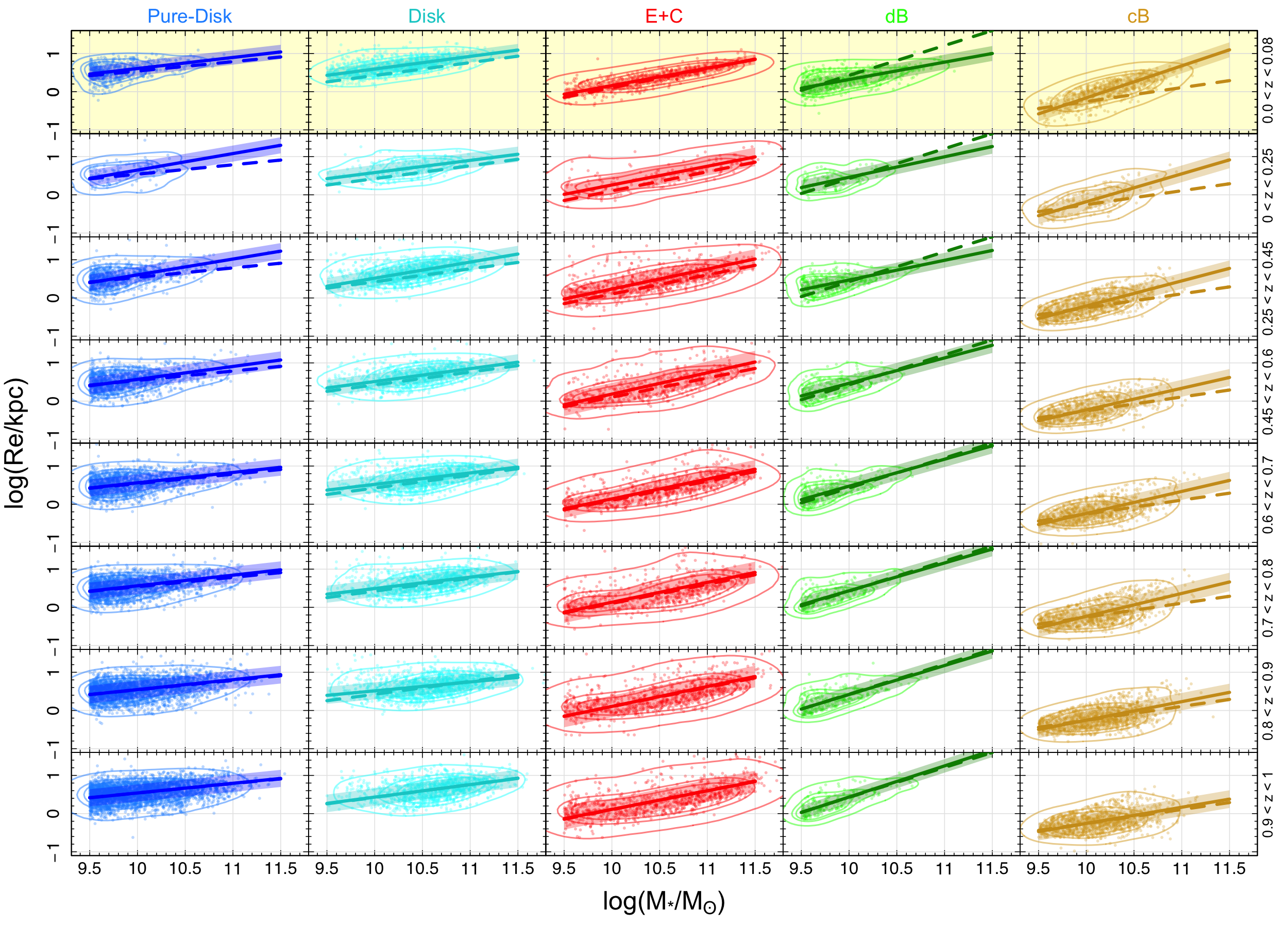}
	\caption{The evolution of the $M_*-R_e$ relations at $0 < z < 1$. Contours represent quantiles showing the levels containing $50\%$, $68\%$ and $95\%$ of the data. Solid lines are fits to the data and dashed lines represent the fit to the data at $z \sim 1$. Transparent regions around lines show the intrinsic scatter along y-axis ($1 \sigma$). Note that the stellar mass on the x-axis represent the component mass not the total mass of host galaxies. The first row highlighted with yellow transparent band shows the redshift range covered by GAMA data (i.e., $0 < z < 0.08$).}
	\label{fig:MRe}
\end{figure}
\end{landscape}

Figure \ref{fig:MRe} shows the $M_*-R_e$ relations of our D10/ACS sample together with $z \sim 0$ GAMA galaxies highlighted with yellow colour. Note that we use the $R_e$ of GAMA galaxies contained within segments as recommended by \cite{Casura-inprep} and stellar masses as measured by \cite{Bellstedt20b}. Likewise, we use the same B/T in r-band (within segments) to calculate the stellar mass of bulges and disks of BD systems, i.e., $M_*^\mathrm{Bulge} = \mathrm{B/T} \times M_*^\mathrm{Total}$ and $M_*^\mathrm{disk} = (1-\mathrm{B/T}) \times M_*^\mathrm{Total}$. 

Following \cite{Lange16} and \cite{Shen03} we adopt a power law function to fit the $M_*-R_e$ relations:

\begin{equation}
    \mathrm{log}(R_e/kpc) = a\,\mathrm{log}(M_*/M_\odot)-b,
    \label{eq:MRe}
\end{equation}

\noindent where $R_e$ is the half-light radius or effective radius in unit of kpc and $M_*$ represents the stellar mass of each component. We then use the {\sc HyperFit} package \citep{Robotham15} to fit Equation \ref{eq:MRe} to the data, adopting a Markov Chain Monte Carlo (MCMC) minimisation method using a Componentwise Hit-And-Run Metropolis (CHARM) algorithm with $10,000$ iterations. See \cite{Robotham15} for more details about {\sc HyperFit}. 

Figure \ref{fig:MRe} shows both the distribution of the data in the $M_*-R_e$ plane and the fit to data. The transparent region around the best fit lines represent the $1 \sigma$ intrinsic scatter along the y-axis. The over-plotted dashed lines are $M_*-R_e$ relations at $z \sim 1$ to highlight the evolution across time.   

In general, the evolution of the $M_*-R_e$ relation shows that disks are larger than all other structures at all epochs while at later times ellipticals start to become larger at the high-mass end. Further, while pure-disks and disk structures follow similar relations, at a given $M_*$ disks located in bulge+disk systems are relatively more massive than pure-disk systems. Figure \ref{fig:MRe} also highlights that at a given mass cBs are smaller and more massive than dBs at all redshifts, implying that cBs are denser structures, which is expected by our dB/cB definitions. 

\begin{figure} 
	\centering
	\includegraphics[width = \columnwidth]{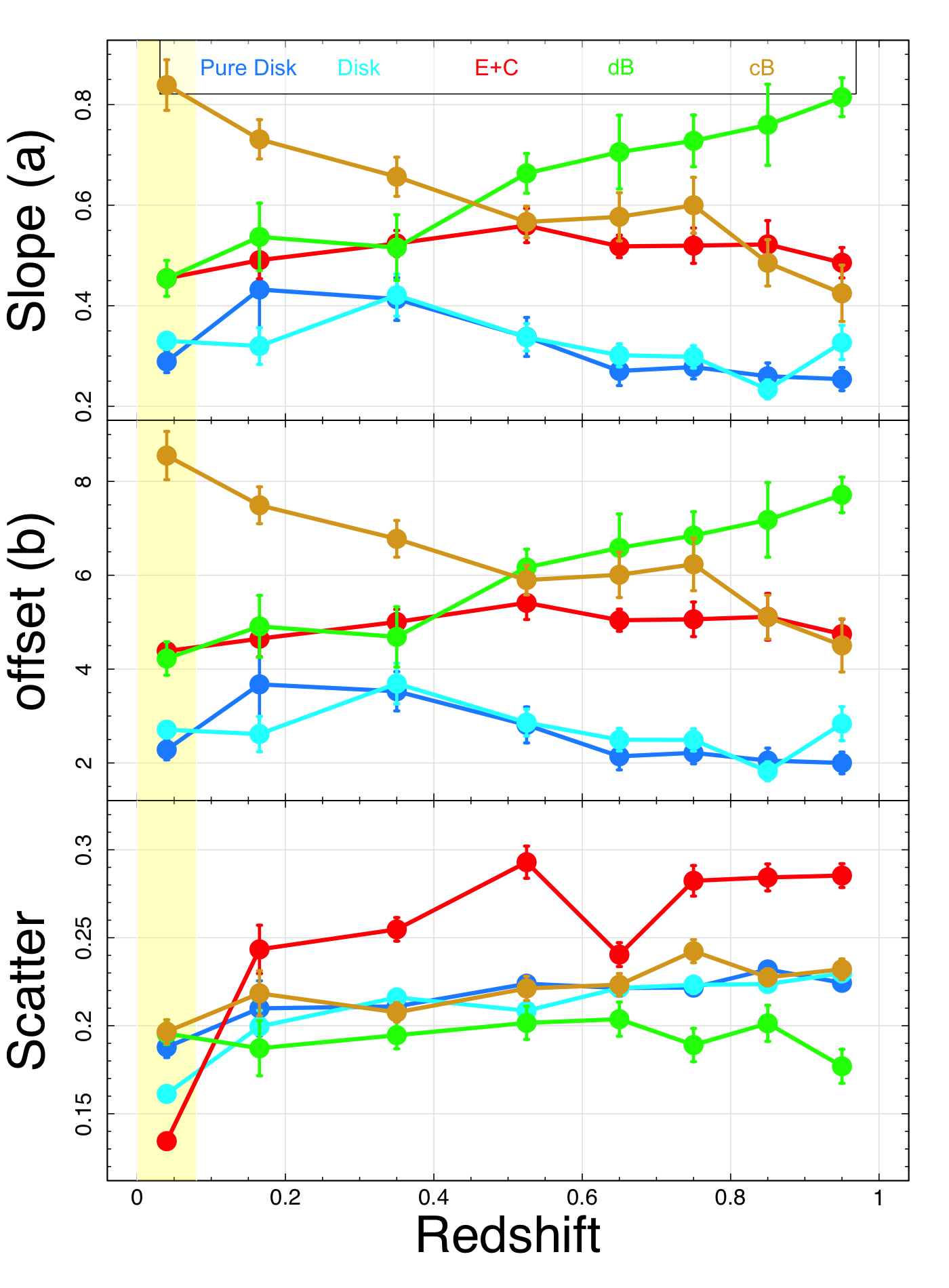}
	\caption{The evolution of the slope (a; top), offset (b; middle) and intrinsic scatter (bottom) of the $M_*-R_e$ relations. The yellow transparent band shows the redshift range covered by GAMA data. }
	\label{fig:MRe_scatter}
\end{figure}

Investigating Figure \ref{fig:MRe}, we find that the $M_*-R_e$ relation of pure-disk systems (blue) and disk components (cyan) are consistent with modest discernible evolution from $z = 1$ to $z \sim 0$. This result is in agreement with other studies for example \cite{Ravindranath04}, \cite{Barden05}, \cite{Mosleh17}.   
We also show this trend in Figure \ref{fig:MRe_scatter} where we plot the evolution of the best fit slope (a), offset (b) and scatter of the $M_*-R_e$ relations and highlight how pure-disk systems follow the same best fit parameters as disk components across all the redshift range suggesting that, as expected, they have similar origins. 

The E+C population experience what is consistent with no evolution in their $M_*-R_e$ relation from $z = 1$ to 0. According to Hashemizadeh et al. (in review), the stellar mass growth in E+C population is dominated by the growth in their high-mass end, indicating that these systems move along their $M_*-R_e$ relation. This is consistent with other studies showing a significant growth of early-type galaxies in size (see e.g., \citealt{Navarro00}; \citealt{Trujillo06}; \citealt{vanderWel08}; \citealt{vanderWel14}).

Figure \ref{fig:MRe_scatter} further summarizes the above results and displays the evolution of the slope (a), offset (b) and intrinsic scatter, $\sigma$, of the $M_*-R_e$ relations (top and bottom panels, respectively).
As shown in the top panel of Figure \ref{fig:MRe_scatter}, dBs (green) have the steepest relation by $z \sim 0.45$ before cBs (gold) become the steepest relation. We note that dBs experience a flattening with cosmic time that is, interestingly, mapped to a significant steepening of cB relation. The top panel of this figure also shows that disks (both pure-disks; blue and disk components; cyan) have flattest relations with lowest offset (middle panel) while Es experience a somewhat unchanged slope ($a \sim 0.5$) and offset ($b \sim 5$).
We note that, despite the evolution and transition between morphological types due for example to mergers, accretions, bulge formation, etc; see \citealt{Hashemizadeh21} and Hashemizadeh et al., in review the $M_*-R_e$ relations are relatively stable and hence well established at $z = 1$. 



We calculate the scatter of the $M_*-R_e$ distributions along the y-axis, i.e., $\mathrm{log}(R_e)$ (bottom panel), and we find that the scatter of all structures are unchanged across redshift (with an average value of $\sim 0.2-0.25$; see Table \ref{tab:MRe_evol} for exact values). At this stage we cannot entirely rule out whether the scatters of our dB/cB $M_*-R_e$ relations could be a by-product of our structural decomposition and/or dB/cB separation. However, the stability of the other component classes would suggest it is intrinsic. Figure \ref{fig:MRe_scatter} also shows that Es have a slightly broader $M_*-R_e$ distribution than other structures and its scatter decreases with time while dBs have the lowest scatter.    

We report the best fit parameters of Equation \ref{eq:MRe} to each structure at different redshift bins in Table \ref{tab:MRe_evol}.

\begin{table*}
\centering
\caption{Best fit parameters of Equation \ref{eq:MRe} at different redshift bins. }
\begin{adjustbox}{scale = 0.8}
\begin{tabular}{lcccccccc}
\firsthline \firsthline \\
$z$-bins      &  $0.0 \leq z < 0.08$  &  $0.0 \leq z < 0.25$ &  $0.25 \leq z < 0.45$ &  $0.45 \leq z < 0.60$ &  $0.60 \leq z < 0.70$ &  $0.70 \leq z < 0.80$ &  $0.80 \leq z < 0.90$ &  $0.90 \leq z \leq 1.00$ \\ \\ \hline                     
                    &   \multicolumn{8}{c}{\textbf{Pure-Disk}}  \\
\cline{2-9} \\
$a$           & $0.364\pm0.048$   & $ 0.403\pm0.086$ & $0.432\pm0.055$ & $0.346\pm0.039$ & $0.318\pm0.041$ & $0.285\pm0.024$ & $0.261\pm0.022$ & $0.257\pm0.026$ \\ \\
$b$           & $3.022\pm0.475$   & $3.391\pm0.843$  & $3.715\pm0.541$ & $2.889\pm0.390$ & $2.619\pm0.405$ & $2.285\pm0.237$ & $2.066\pm0.224$ & $2.034\pm0.263$ \\ \\
scatter       & $0.190\pm0.006$   & $0.207\pm0.015$  & $0.212\pm0.006$ & $0.224\pm0.005$ & $0.223\pm0.004$ & $0.222\pm0.004$ & $0.232\pm0.003$ & $0.225\pm0.003$ \\ \\ \hline 
                    &   \multicolumn{8}{c}{\textbf{Disk Component}}  \\
\cline{2-9} \\
$a$         & $0.327\pm0.016$ & $0.354\pm0.034$  & $0.390\pm0.029$ & $0.333\pm0.023$ & $0.327\pm0.034$ & $0.332\pm0.040$ & $0.253\pm0.038$ & $0.330\pm0.043$  \\ \\
$b$         & $2.674\pm0.165$ & $2.968\pm0.353$  & $3.366\pm0.300$ & $2.811\pm0.239$ & $2.768\pm0.353$ & $2.849\pm0.418$ & $2.024\pm0.406$ & $2.870\pm0.459$  \\ \\
scatter     & $0.161\pm0.003$ & $0.201\pm0.009$  & $0.215\pm0.005$ & $0.208\pm0.005$ & $0.223\pm0.005$ & $0.225\pm0.005$ & $0.225\pm0.004$ & $0.230\pm0.005$  \\ \\ \hline 
                    &   \multicolumn{8}{c}{\textbf{Diffuse-Bulge}}  \\
\cline{2-9} \\ 
$a$          & $0.447\pm0.038$  & $0.521\pm0.078$ & $0.481\pm0.033$ & $0.726\pm0.095$ & $0.710\pm0.048$ & $0.722\pm0.054$ & $0.744\pm0.060$ & $0.849\pm0.061$ \\ \\
$b$          & $4.147\pm0.377$  & $4.756\pm0.770$ & $4.347\pm0.324$ & $6.786\pm0.933$ & $6.628\pm0.467$ & $6.785\pm0.536$ & $7.027\pm0.600$ & $8.044\pm0.599$  \\\\
scatter      & $0.195\pm0.006$  & $0.186\pm0.015$ & $0.193\pm0.008$ & $0.205\pm0.011$ & $0.204\pm0.009$ & $0.189\pm0.010$ & $0.200\pm0.010$ & $0.179\pm0.010$ \\ \\\hline 
                    &   \multicolumn{8}{c}{\textbf{Compact-Bulge}}  \\
\cline{2-9} \\ 
$a$           & $0.820\pm0.036$  & $0.779\pm0.076$ & $0.677\pm0.046$ & $0.599\pm0.044$ & $0.558\pm0.047$ & $0.647\pm0.066$ & $0.498\pm0.055$ & $0.405\pm0.042$ \\ \\
$b$           & $8.355\pm0.369$  & $7.972\pm0.765$ & $6.981\pm0.465$ & $6.224\pm0.445$ & $5.822\pm0.470$ & $6.708\pm0.666$ & $5.238\pm0.555$ & $4.306\pm0.426$  \\ \\ 
scatter       & $0.195\pm0.006$  & $0.221\pm0.013$ & $0.209\pm0.005$ & $0.223\pm0.007$ & $0.223\pm0.006$ & $0.245\pm0.007$ & $0.228\pm0.005$ & $0.231\pm0.005$ \\ \\ \hline 
                    &   \multicolumn{8}{c}{\textbf{Elliptical+Compact}}  \\
\cline{2-9} \\
$a$           & $0.440\pm0.006$   & $0.471\pm0.025$ & $ 0.480\pm0.018$ & $0.574\pm0.043$ & $0.533\pm0.034$ & $0.492\pm0.030$ & $0.505\pm0.035$ & $0.496\pm0.034$ \\ \\
$b$           & $4.227\pm0.068$   & $4.446\pm0.266$ & $4.547\pm0.193$ & $5.560\pm0.451$  & $5.204\pm0.354$ & $4.776\pm0.313$ & $4.939\pm0.373$ & $4.856\pm0.355$  \\ \\ 
scatter       & $0.134\pm0.004$   & $0.240\pm0.013$ & $0.252\pm0.006$ & $0.295\pm0.010$  & $0.241\pm0.007$ & $0.280\pm0.008$ & $0.282\pm0.007$ & $0.286\pm0.007$ \\ \\ \hline 

\lasthline
\end{tabular}
\end{adjustbox}
\label{tab:MRe_evol}
\end{table*}

\section{Comparison with the literature}
\label{sec:literature_comp}

\begin{figure*} 
	\centering
	\includegraphics[width = 0.8\textwidth]{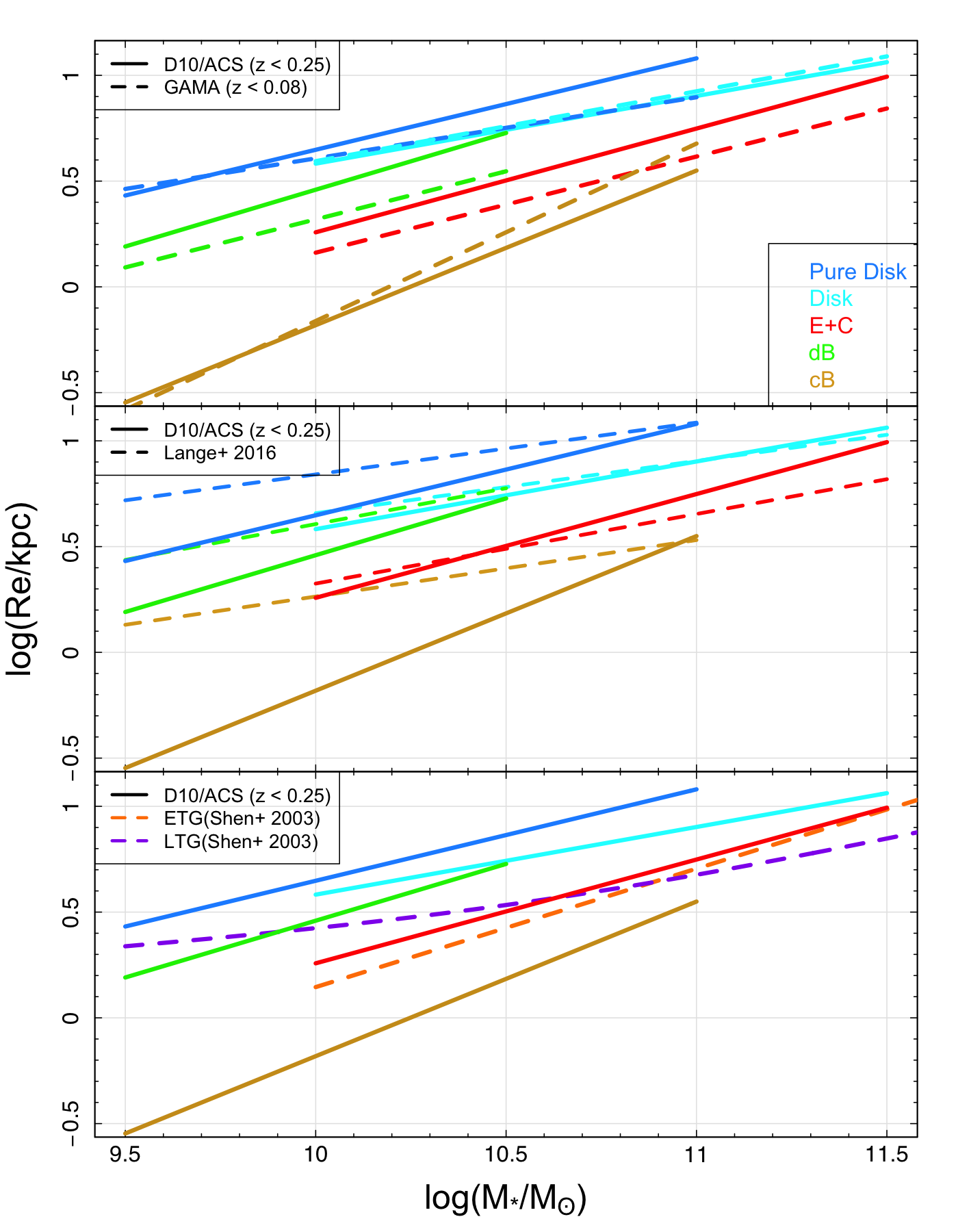}
	\caption{Comparison of our GAMA and DEVILS low-$z$ $M_*-R_e$ relations (left and right panels, respectively) compared with \protect\cite{Lange16} and \protect\cite{Shen03} measurements (dashed and dotted lines, respectively). Note that dashed green and gold lines represent late- and early-type bulges in \protect\cite{Lange16} that might not be comparable to our dB and cB classification. }
	\label{fig:MRe_Lange_Shen}
\end{figure*}

In this section, we compare our $M_*-R_e$ relations with key literature results at both low and high redshifts. 
Figure \ref{fig:MRe_Lange_Shen} shows the comparison of our D10/ACS low-$z$ ($z < 0.25$) $M_*-R_e$ relations with our GAMA sample (\citealt{Casura-inprep}) as well as with previous works by \cite{Lange16} for GAMA galaxies and \cite{Shen03} for SDSS galaxies. 

As shown in the top panel of Figure \ref{fig:MRe_Lange_Shen}, we find that despite the difference in redshift ranges ($z < 0.25$ for the D10/ACS versus $z < 0.08$ for GAMA) the $M_*-R_e$ relation of both pure-disk population and disk components of the D10/ACS are in good agreement with the GAMA relations. We find that at a given stellar mass the D10/ACS elliptical galaxies are larger in size than those of the GAMA sample, although they both follow relatively similar slope. The relation of our cB population is also consistent with GAMA while dB population is slightly more contracted in GAMA than in D10/ACS. We note again that we are comparing two very distinct types of data with significant difference in resolution, PSF etc. Moreover, the structural analysis of GAMA and D10/ACS have been done by different groups with different settings and pipelines. Therefore, one might naturally expect some mismatch between these two data-sets. These effects remain under investigation, and such work is outside the scope of this paper.     

We can also compare our D10/ACS low-$z$ $M_*-R_e$ relations with the earlier measurements of GAMA galaxies presented by \cite{Lange16} (shown in the middle panel of Figure \ref{fig:MRe_Lange_Shen}). The structural divisions of \cite{Lange16} are not directly comparable with our classification, however, for completeness, we compare our pure-disk, disk component, dB, cB and E+C systems with their late-type disk, early-type disk, late-type bulge, early-type bulge and elliptical, respectively. We find that our $M_*-R_e$ relations of disk components and ellipticals are in good agreement with \cite{Lange16}. However, the relation for other structures are inconsistent likely due to our different classification techniques than that in \cite{Lange16}. This difference is more obvious in bulge structures where, as expected, morphological classification and structural analysis are important. This large difference in bulge relations (both cBs and dBs) are expected as with the HST high-resolution imaging we are able to resolve much smaller bulges at this low redshift range leading to steeper $M_*-R_e$ relations.

In the lower panel of Figure \ref{fig:MRe_Lange_Shen}) we compare our low-$z$ D10/ACS $M_*-R_e$ relations with the \cite{Shen03} relation for early- and late-type galaxies (ETG and LTG, respectively) and based on SDSS data. Their ETG/LTG classification is based on S\'ersic index (n) with $n < 2.5$ and $n > 2.5$ representing LTGs and ETGs, respectively. Again, their broad distinction is not comparable with our morphological classifications, but we find a relatively good agreement between their ETGs and our Es.    

\begin{figure*} 
	\centering
	\includegraphics[width = \textwidth]{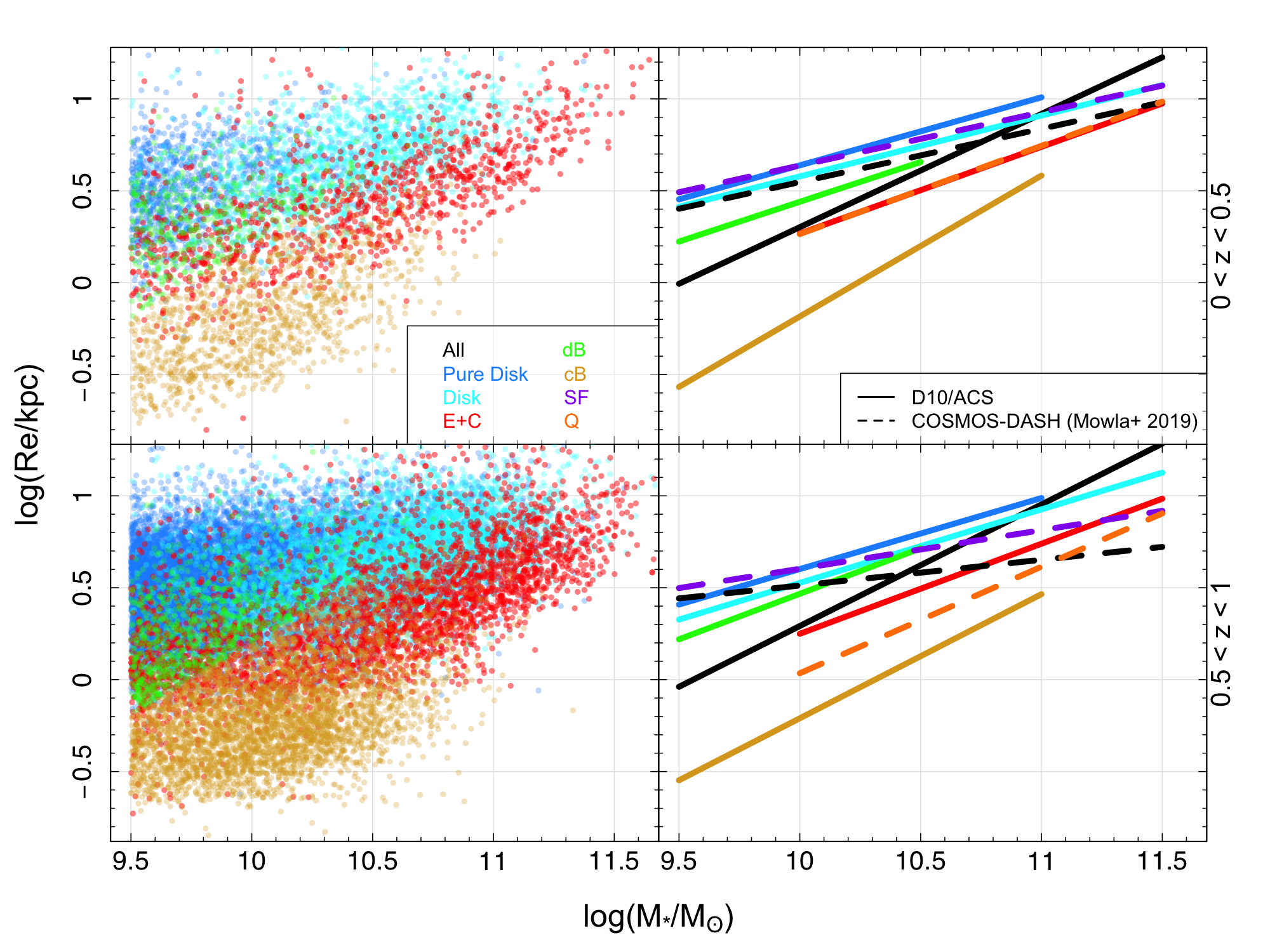}
	\caption{Comparison of our $M_*-R_e$ relations (solid lines) in two redshift ranges of $0.0 < z < 0.5$ and $0.5 < z < 1$ with that of COSMOS-DASH (dashed lines) from \protect\cite{Mowla19}. COSMOS-DASH includes the $M_*-R_e$ relations for star-forming (purple) and quiescent (orange) galaxies.}
	\label{fig:MRe_COMOS_DASH}
\end{figure*}

Finally, in Figure \ref{fig:MRe_COMOS_DASH}, we compare our D10/ACS $M_*-R_e$ relations with higher-$z$ relations from \cite{Mowla19} obtained from a combination of COSMOS-DASH data and previous \cite{vanderWel14} data in CANDELS. \cite{Mowla19} separate their sample into two classes of star-forming (SF) and quiescent (Q) galaxies which is not the same as our morphological categories but we show this comparison for completeness. Here, similar to \cite{Mowla19}, we split our sample into two broad redshift bins of $0.0 < z < 0.5$ and $0.5 < z < 1$ and highlight that their total $M_*-R_e$ relations in both redshift bins are flatter than our D10/ACS. The relation of our E+C population is in excellent agreement with their quiescent galaxies (orange dashed lines) in $0 < z < 0.5$ while offsets in $0.5 < z < 1$. We find that their relation for star-forming systems (purple dashed lines) are consistent with our disk populations.

\section{Size evolution of structures}
\label{sec:size_evol}

\begin{figure*} 
	\centering
	\includegraphics[width = 0.7\textwidth]{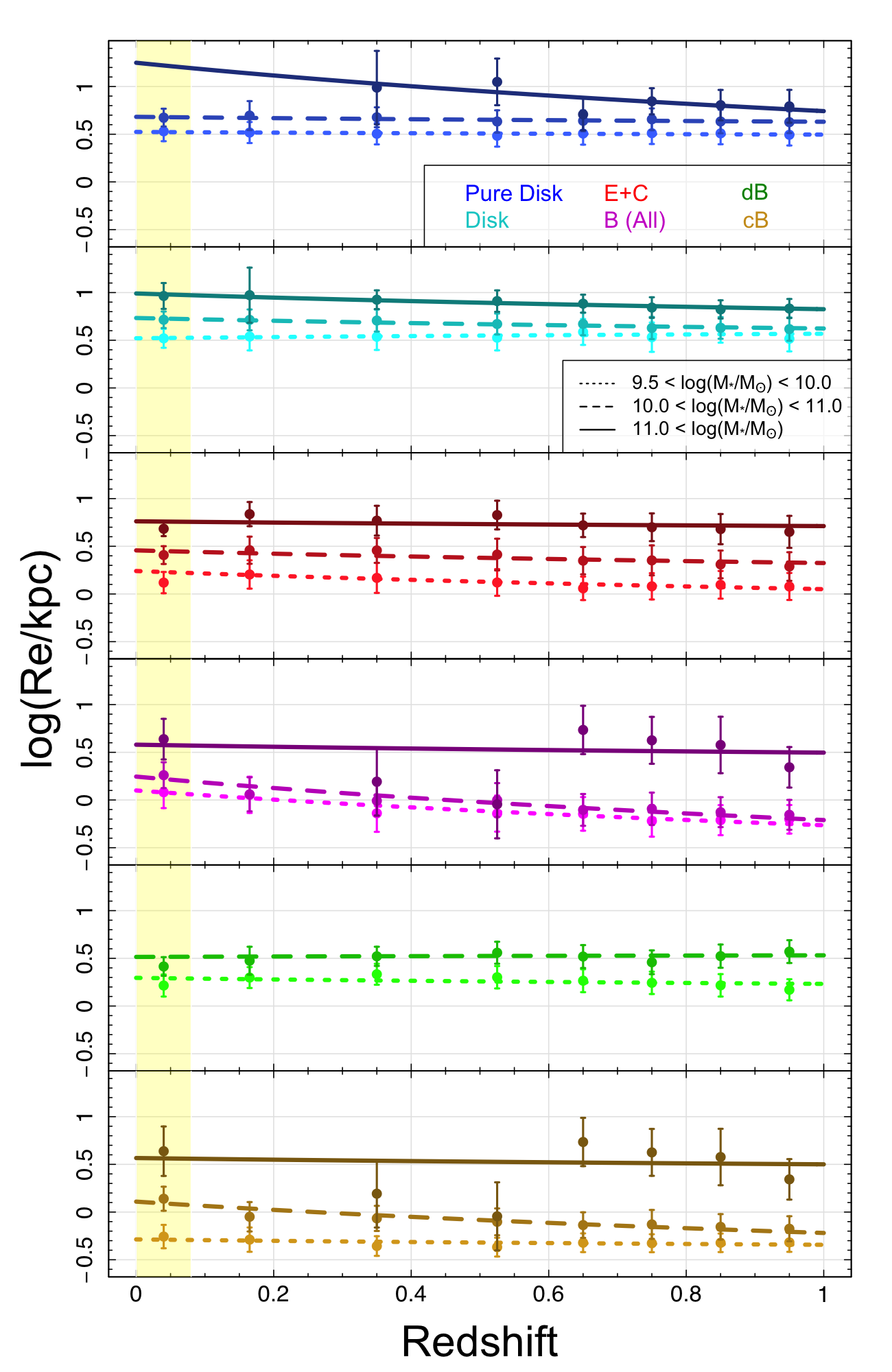}
	\caption{The size evolution of structures since $z = 1$ in three stellar mass bins shown as dotted ($9.5 \leq \mathrm{log}(M_*/M_\odot) < 10.0$,), dashed ($10.0 \leq \mathrm{log}(M_*/M_\odot) < 11.0$) and solid ($11.0 \leq \mathrm{log}(M_*/M_\odot)$) lines. Note that the stellar mass here is the component stellar mass not the total. Data points represent median sizes in each $z$-bin while lines show the results of our parameterized fit of Equation \ref{eq:Re_z} to data points. See Table \ref{tab:Re_z} for best fit parameters.  }
	\label{fig:Re_z}
\end{figure*}

With the mass-size measurements of our D10/ACS sample in place, we now explore the variation of the size of each of the above structures over time. Figure \ref{fig:Re_z} shows this evolution with data points representing the median size of each structure per redshift bin per stellar mass bin. We bin our structures into three bins of component stellar mass, as: $9.5 \leq \mathrm{log}(M_*/M_\odot) < 10.0$, $10.0 \leq \mathrm{log}(M_*/M_\odot) < 11.0$ and  $11.0 \leq \mathrm{log}(M_*/M_\odot)$. 

Following the literature (see e.g., \citealt{vanderWel14}; \citealt{Mosleh17}; \citealt{PaulinoAfonso17}; \citealt{Marshall19}) we fit the size evolution as a function of $z$ with the following function:  

\begin{equation}
    R_e/kpc = \alpha\,(1+z)^\beta,
    \label{eq:Re_z}
\end{equation}

\noindent where $R_e$ is the effective radius in units of kpc and $z$ is redshift. Table \ref{tab:Re_z} summarizes the best fit parameters of Equation \ref{eq:Re_z} to our data. We use a Levenberg-Marquardt nonlinear least-squares
algorithm implemented in the \texttt{minpack.lm} package in {\sc R} for fitting the above equation to our data.

Figure \ref{fig:Re_z} shows that regardless of the type of structure, as expected, more massive components are larger in size than less massive ones throughout the redshift range. For example, the high-mass Es (red) are on average $\sim 3$ times larger than the low-mass ones while high-mass disk components (cyan) are on average $\sim 2$ times larger than low-mass ones. This dependence of mass and size of structures is obviously the direct reflection of the slope of the $M_*-R_e$ relations that we discussed in Section \ref{sec:DEVILS_MRe}, implying that the steeper the relation, the more mass dependent the size is.  

Interestingly, as shown in Figure \ref{fig:Re_z}, we find a minimal size evolution since $z = 1$ for all structures. This unchanged trend is still valid for structures of different stellar mass. 
The best fit parameters reported in Table \ref{tab:Re_z} show that although the growth in size is small, the slope of our parameterized fits, $\beta$, seems to have some correlation with the stellar mass with more massive structures growing slightly faster than lower mass ones (\citealt{Robotham14}). For example, the disk component (cyan) has $\beta \sim -0.16 \pm 0.05$ in their low stellar mass bin versus $\beta \sim -0.55\pm0.07$ in their high-mass range. 
We note that as expected we do not find any dB in the largest stellar mass bin ($\mathrm{log}(M_*/M_\odot) > 11$) while we find only 14 cB systems in this mass regime. Due to this small number of objects we have large errors on data and hence not clear whether the trend is real. In spite of the large uncertainties, this trend is consistent with minimal size evolution.  
However, we do not rule out some effects of our dB/cB distinction procedure on this trend. We also show the size evolution of all bulges (magenta) showing that lower mass bulges, in general, grow by a factor of $\sim 7$ in size since $z = 1$.

\begin{table*}
\centering
\caption{Best fit parameters of Equation \ref{eq:Re_z} for different bins of component stellar mass. Note the we don't find any dB in the last mass bin. }
\begin{adjustbox}{scale = 1}
\begin{tabular}{lccc}
\firsthline \firsthline \\
mass bins      &  $9.0 \leq \mathrm{log}(M_*/M_\odot) < 10.0$  &  $10.0 \leq \mathrm{log}(M_*/M_\odot) < 11.0$ &  $11.0 \leq \mathrm{log}(M_*/M_\odot)$ \\ \\ \hline                
                    &   \multicolumn{3}{c}{\textbf{Pure-Disk}}  \\
\cline{2-4} \\
$\alpha$           & $3.342\pm0.086$    & $4.811\pm0.126$ & $17.753\pm7.366$  \\ \\
$\beta$            & $-0.093\pm0.054$   & $-0.171\pm0.056$ & $-1.684\pm0.866$  \\ \\ \hline
                   &   \multicolumn{3}{c}{\textbf{Disk Component}}  \\
\cline{2-4} \\
$\alpha$           & $3.318\pm0.233$    & $5.411\pm0.123$  & $9.783\pm0.295$  \\ \\
$\beta$            & $0.157\pm0.154$    & $-0.365\pm0.054$ & $-0.547\pm0.067$  \\ \\ \hline
                   &   \multicolumn{3}{c}{\textbf{Bulge (All)}}  \\
\cline{2-4} \\
$\alpha$           & $1.259\pm0.070$    & $1.759\pm0.128$  & $3.802\pm1.536$  \\ \\
$\beta$            & $-1.211\pm0.149$   & $-1.513\pm0.219$ & $-0.276\pm0.854$  \\ \\ \hline
                   &   \multicolumn{3}{c}{\textbf{Diffuse-Bulge}}  \\
\cline{2-4} \\
$\alpha$           & $1.974\pm0.178$    & $3.272\pm0.528$ & $-$  \\ \\
$\beta$            & $-0.209\pm0.204$   & $0.056\pm0.310$ & $-$  \\ \\ \hline
                   &   \multicolumn{3}{c}{\textbf{Compact-Bulge}}  \\
\cline{2-4} \\
$\alpha$           & $0.517\pm0.030$    & $1.286\pm0.095$   & $3.685\pm1.648$  \\ \\
$\beta$            & $-0.188\pm0.125$   & $-1.091\pm0.198$  & $-0.219\pm0.919$  \\ \\ \hline
                   &   \multicolumn{3}{c}{\textbf{Elliptical+Compact}}  \\
\cline{2-4} \\
$\alpha$           & $1.740\pm0.097$    & $2.864\pm0.181$  & $5.778\pm0.638$  \\ \\
$\beta$            & $-0.633\pm0.121$   & $-0.440\pm0.156$ & $-0.166\pm0.266$  \\ \\ 
\lasthline
\end{tabular}
\end{adjustbox}
\label{tab:Re_z}
\end{table*}

\section{Summary and conclusion}

In this work, using our structural analysis, we have investigated the evolution of the stellar mass-size relation since $z = 1$. We find close-to-no-variation in the $M_*-R_e$ relation for almost all structures indicating that in spite of ongoing although declining star-formation, mass evolution, morphological transitions and mergers the $M_*-R_e$ relations only modestly varies. This implies that in general evolution moves galaxies along their $M_*-R_e$ relations. Although the results of Paper-II shows a fairly moderate mass movement. 

We find that in agreement with other studies (e.g., \citealt{Trujillo06}; \citealt{Lange16}) our E+C class and bulge components follow steeper $M_*-R_e$ relations than disk structures indicating that at a given stellar mass Es and bulges are smaller than disk structures but also that they likely build up via distinct evolutionary mechanisms. Note that this is valid until $\mathrm{M_*}/\mathrm{M}_\odot < 10^{11}$, while more massive ellipticals are on average larger than disks and potentially the largest structures. These are likely cDs in clusters cores.

Our results show that the slope of the $M_*-R_e$ relations of pure disk systems at all redshift ranges is very consistent with that of disk components of bulge+disk systems (see Figures \ref{fig:MRe} and \ref{fig:MRe_scatter}) suggesting the same origin and evolutionary pathway for all disks regardless of the presence or absence of a bulge component.
In addition, in agreement with other studies (e.g., \citealt{Ravindranath04} and \citealt{Barden05}), we find a redshift independent $M_*-R_e$ relation for disks with a slope of $\sim 0.3$. This is also in agreement with several studies that in a broader classification have shown that the $M_*-R_e$ of late-type galaxies evolves only very little in $z < 1$ (see e.g., \citealt{Lilly98}; \citealt{Dutton11}; \citealt{vanderWel14}; \citealt{Mosleh17}).  

Exploring the variation of the $M_*-R_e$ relation of the E+C galaxies we find that this relation experience a what is consistent with no change while we have shown in Paper-II (\citealt{Hashemizadeh22}) that the majority of the evolution in stellar mass happens at their high mass end (see Figures \ref{fig:MRe} and \ref{fig:MRe_scatter}). Note, however, that Figure \ref{fig:Re_z} shows that low-mass E+C systems seem to grow slightly faster in size than high-mass ones. 
This further confirms our results in Paper-I and II suggesting that, in agreement with \cite{Robotham14}, at $z < 1$ minor mergers are the dominant driver of the growth/formation of E systems. This interpretation is in agreement with other observational (e.g., \citealt{Trujillo11}) and theoretical (e.g., \citealt{Naab09}; \citealt{Xie15}) studies identifying minor mergers to play the most important role in the growth of E galaxies. 

In addition, compact bulges also present a steepening with time, whilst diffuse-bulges experience only a modest flattening (contraction). We find different $M_*-R_e$ relation for cBs than that of Es likely suggesting that they have different origins. This further highlights our argument that in-situ and secular star formation are the dominant processes in bulge formation in the $z < 1$ Universe given that we concluded in Paper-II that dBs have the largest growth rate in their stellar mass density.

We also investigated the size evolution of structures (Figure \ref{fig:Re_z}) and concluded that as expected more massive structures are larger in size, too, with only a modest variation since $z = 1$. We showed that although disk structures grow a little in size, bulges (of $\mathrm{log}(M_*/M_\odot) < 11$, in particular) grow slightly and eventually by a factor of $\sim 3$ at $z \sim 0$.  

In summary, we find a surprisingly unchanged size and $M_*-R_e$ relation over the last $\sim 8$ Gyr. This lack of evolution particularly in disk structures is consistent with our previous results, suggesting that evolution is predominantly along the respective $M_*-R_e$ relations. In effect, the scaling relations lay down the trail along which galaxies have evolved. Each trail hence requiring a distinct path, i.e. an inside-out growth and evolution in which the size of galaxies grows as they grow in stellar mass. In addition, minor mergers seem to be responsible for the growth of elliptical systems at least since $z < 1$. However, further investigations, numerical simulations in particular, is required to confirm these results and put them into a comprehensive physical picture. 
      
\section*{Acknowledgements}
DEVILS is an Australian project based around a spectroscopic campaign using the Anglo-Australian Telescope. The DEVILS input catalogue is generated from data taken as part of the ESO VISTA-VIDEO \citep{Jarvis13} and UltraVISTA \citep{McCracken12} surveys. DEVILS is part funded via Discovery Programs by the Australian Research Council and the participating institutions. The DEVILS website is \href{https://devilsurvey.org}{https://devilsurvey.org}. The DEVILS data is hosted and provided by AAO Data Central (\href{https://datacentral.org.au}{https://datacentral.org.au}). This work was supported by resources provided by The Pawsey Supercomputing Centre with funding from the Australian Government and the Government of Western Australia. We also gratefully acknowledge the contribution of the entire COSMOS collaboration consisting of more than 200 scientists. The HST COSMOS Treasury program was supported through NASA grant HST-GO-09822. SB and SPD acknowledge support by the Australian Research Council's funding scheme DP180103740. MS has been supported by the European Union's  Horizon 2020 research and innovation programme under the Maria Skłodowska-Curie (grant agreement No 754510), the National Science Centre of Poland (grant UMO-2016/23/N/ST9/02963) and by the Spanish Ministry of Science and Innovation through Juan de la Cierva-formacion program (reference FJC2018-038792-I). ASGR and LJMD acknowledge support from the Australian Research Council's Future Fellowship scheme (FT200100375 and FT200100055, respectively).

This work was made possible by the free and open R software (\citealt{R-Core-Team}).
A number of figures in this paper were generated using the R \texttt{magicaxis} package (\citealt{Robotham16b}). This work also makes use of the \texttt{celestial} package (\citealt{Robotham16a}) and \texttt{dftools} (\citealt{Obreschkow18}).

\subsection{Data Availability} 
\label{subsec:data_access}

The catalogues used in this paper are \texttt{D10VisualMoprhologyCat}, described in \cite{Hashemizadeh21}, and \texttt{DEVILS\_BD\_Decomp} and is held on the DEVILS database managed by AAO Data Central (\href{https://datacentral.org.au}{https://datacentral.org.au}). The catalogues are currently only available for the DEVILS team, but will be made publicly available in a future DEVILS data release.  All imaging data are in the public domain and were downloaded from the the NASA/IPAC Infrared Science Archive (IRSA) web-page: \href{https://irsa.ipac.caltech.edu/data/COSMOS/images/acs\_2.0/I/}{irsa.ipac.caltech.edu/data/COSMOS/images/acs\_2.0/I/}. The main tools used in this study are {\sc ProFit} \citep{Robotham17} version 1.3.3 (available at: \href{https://github.com/ICRAR/ProFit}{https://github.com/ICRAR/ProFit}) and ProFound \citep{Robotham18} version 1.3.4 (available at: \href{https://github.com/asgr/ProFound}{https://github.com/asgr/ProFound}). We used {\sc Tiny Tim} version 6.3 to generate the HST/ACS Point Spread Function (PSF). We further use {\sc LaplacesDemon} version 1.3.4 implemented in {\sc R} available at: \href{https://github.com/LaplacesDemonR/LaplacesDemon}{https://github.com/LaplacesDemonR/LaplacesDemon}. Our structural decomposition pipeline, {\sc GRAFit}, is available at: \href{https://github.com/HoseinHashemi/GRAFit}{https://github.com/HoseinHashemi/GRAFit}.

\bibliographystyle{mnras}
\bibliography{library}

\newcommand{\noop}[1]{}
\begin{thebibliography}{}
\makeatletter
\relax
\def\mn@urlcharsother{\let\do\@makeother \do\$\do\&\do\#\do\^\do\_\do\%\do\~}
\def\mn@doi{\begingroup\mn@urlcharsother \@ifnextchar [ {\mn@doi@}
  {\mn@doi@[]}}
\def\mn@doi@[#1]#2{\def\@tempa{#1}\ifx\@tempa\@empty \href
  {http://dx.doi.org/#2} {doi:#2}\else \href {http://dx.doi.org/#2} {#1}\fi
  \endgroup}
\def\mn@eprint#1#2{\mn@eprint@#1:#2::\@nil}
\def\mn@eprint@arXiv#1{\href {http://arxiv.org/abs/#1} {{\tt arXiv:#1}}}
\def\mn@eprint@dblp#1{\href {http://dblp.uni-trier.de/rec/bibtex/#1.xml}
  {dblp:#1}}
\def\mn@eprint@#1:#2:#3:#4\@nil{\def\@tempa {#1}\def\@tempb {#2}\def\@tempc
  {#3}\ifx \@tempc \@empty \let \@tempc \@tempb \let \@tempb \@tempa \fi \ifx
  \@tempb \@empty \def\@tempb {arXiv}\fi \@ifundefined
  {mn@eprint@\@tempb}{\@tempb:\@tempc}{\expandafter \expandafter \csname
  mn@eprint@\@tempb\endcsname \expandafter{\@tempc}}}

\bibitem[\protect\citeauthoryear{{Barden} et~al.,}{{Barden}
  et~al.}{2005}]{Barden05}
{Barden} M.,  et~al., 2005, \mn@doi [\apj] {10.1086/497679}, \href
  {https://ui.adsabs.harvard.edu/abs/2005ApJ...635..959B} {635, 959}

\bibitem[\protect\citeauthoryear{{Bellstedt} et~al.,}{{Bellstedt}
  et~al.}{2020}]{Bellstedt20b}
{Bellstedt} S.,  et~al., 2020, arXiv e-prints, \href
  {https://ui.adsabs.harvard.edu/abs/2020arXiv200511917B} {p. arXiv:2005.11917}

\bibitem[\protect\citeauthoryear{{Bouwens}, {Cay{\'o}n}  \& {Silk}}{{Bouwens}
  et~al.}{1997}]{Bouwens97}
{Bouwens} R.~J.,  {Cay{\'o}n} L.,   {Silk} J.,  1997, \mn@doi [\apjl]
  {10.1086/310969}, \href
  {https://ui.adsabs.harvard.edu/abs/1997ApJ...489L..21B} {489, L21}

\bibitem[\protect\citeauthoryear{{Casura} et~al.,}{{Casura}
  et~al.}{prep}]{Casura-inprep}
{Casura} S.,  et~al., in-prep

\bibitem[\protect\citeauthoryear{{Cayon}, {Silk}  \& {Charlot}}{{Cayon}
  et~al.}{1996}]{Cayon96}
{Cayon} L.,  {Silk} J.,   {Charlot} S.,  1996, \mn@doi [\apjl]
  {10.1086/310202}, \href
  {https://ui.adsabs.harvard.edu/abs/1996ApJ...467L..53C} {467, L53}

\bibitem[\protect\citeauthoryear{{Dalcanton}, {Spergel}  \&
  {Summers}}{{Dalcanton} et~al.}{1997}]{Dalcanton97}
{Dalcanton} J.~J.,  {Spergel} D.~N.,   {Summers} F.~J.,  1997, \mn@doi [\apj]
  {10.1086/304182}, \href
  {https://ui.adsabs.harvard.edu/abs/1997ApJ...482..659D} {482, 659}

\bibitem[\protect\citeauthoryear{{Driver} et~al.,}{{Driver}
  et~al.}{2011}]{Driver11}
{Driver} S.~P.,  et~al., 2011, \mn@doi [\mnras]
  {10.1111/j.1365-2966.2010.18188.x}, \href
  {https://ui.adsabs.harvard.edu/abs/2011MNRAS.413..971D} {413, 971}

\bibitem[\protect\citeauthoryear{{Dutton} et~al.,}{{Dutton}
  et~al.}{2011}]{Dutton11}
{Dutton} A.~A.,  et~al., 2011, \mn@doi [\mnras]
  {10.1111/j.1365-2966.2010.17555.x}, \href
  {https://ui.adsabs.harvard.edu/abs/2011MNRAS.410.1660D} {410, 1660}

\bibitem[\protect\citeauthoryear{{Freeman}}{{Freeman}}{1970}]{Freeman70}
{Freeman} K.~C.,  1970, \mn@doi [\apj] {10.1086/150474}, \href
  {https://ui.adsabs.harvard.edu/abs/1970ApJ...160..811F} {160, 811}

\bibitem[\protect\citeauthoryear{{Grogin} et~al.,}{{Grogin}
  et~al.}{2011}]{Grogin11}
{Grogin} N.~A.,  et~al., 2011, \mn@doi [\apjs] {10.1088/0067-0049/197/2/35},
  \href {https://ui.adsabs.harvard.edu/abs/2011ApJS..197...35G} {197, 35}

\bibitem[\protect\citeauthoryear{{Hashemizadeh} et~al.,}{{Hashemizadeh}
  et~al.}{2021}]{Hashemizadeh21}
{Hashemizadeh} A.,  et~al., 2021, \mn@doi [\mnras] {10.1093/mnras/stab600},
  \href {https://ui.adsabs.harvard.edu/abs/2021MNRAS.505..136H} {505, 136}

\bibitem[\protect\citeauthoryear{{Hashemizadeh} et~al.,}{{Hashemizadeh}
  et~al.}{2022}]{Hashemizadeh22}
{Hashemizadeh} A.,  et~al., 2022, arXiv e-prints, \href
  {https://ui.adsabs.harvard.edu/abs/2022arXiv220300185H} {p. arXiv:2203.00185}

\bibitem[\protect\citeauthoryear{{Jarvis} et~al.,}{{Jarvis}
  et~al.}{2013}]{Jarvis13}
{Jarvis} M.~J.,  et~al., 2013, \mn@doi [\mnras] {10.1093/mnras/sts118}, \href
  {https://ui.adsabs.harvard.edu/abs/2013MNRAS.428.1281J} {428, 1281}

\bibitem[\protect\citeauthoryear{{Jiang} et~al.,}{{Jiang}
  et~al.}{2019}]{Jiang19}
{Jiang} F.,  et~al., 2019, \mn@doi [\mnras] {10.1093/mnras/stz1952}, \href
  {https://ui.adsabs.harvard.edu/abs/2019MNRAS.488.4801J} {488, 4801}

\bibitem[\protect\citeauthoryear{{Kauffmann} et~al.,}{{Kauffmann}
  et~al.}{2003}]{Kauffmann03}
{Kauffmann} G.,  et~al., 2003, \mn@doi [\mnras]
  {10.1046/j.1365-8711.2003.06291.x}, \href
  {https://ui.adsabs.harvard.edu/abs/2003MNRAS.341...33K} {341, 33}

\bibitem[\protect\citeauthoryear{{Koekemoer} et~al.,}{{Koekemoer}
  et~al.}{2011}]{Koekemoer11}
{Koekemoer} A.~M.,  et~al., 2011, \mn@doi [\apjs] {10.1088/0067-0049/197/2/36},
  \href {https://ui.adsabs.harvard.edu/abs/2011ApJS..197...36K} {197, 36}

\bibitem[\protect\citeauthoryear{{Kormendy} \& {Norman}}{{Kormendy} \&
  {Norman}}{1979}]{Kormendy79}
{Kormendy} J.,  {Norman} C.~A.,  1979, \mn@doi [\apj] {10.1086/157414}, \href
  {https://ui.adsabs.harvard.edu/abs/1979ApJ...233..539K} {233, 539}

\bibitem[\protect\citeauthoryear{{Lange} et~al.,}{{Lange}
  et~al.}{2016}]{Lange16}
{Lange} R.,  et~al., 2016, \mn@doi [\mnras] {10.1093/mnras/stw1495}, \href
  {https://ui.adsabs.harvard.edu/abs/2016MNRAS.462.1470L} {462, 1470}

\bibitem[\protect\citeauthoryear{{Lilly} et~al.,}{{Lilly}
  et~al.}{1998}]{Lilly98}
{Lilly} S.,  et~al., 1998, \mn@doi [\apj] {10.1086/305713}, \href
  {https://ui.adsabs.harvard.edu/abs/1998ApJ...500...75L} {500, 75}

\bibitem[\protect\citeauthoryear{{Marshall}, {Mutch}, {Qin}, {Poole}  \&
  {Wyithe}}{{Marshall} et~al.}{2019}]{Marshall19}
{Marshall} M.~A.,  {Mutch} S.~J.,  {Qin} Y.,  {Poole} G.~B.,   {Wyithe} J.
  S.~B.,  2019, \mn@doi [\mnras] {10.1093/mnras/stz1810}, \href
  {https://ui.adsabs.harvard.edu/abs/2019MNRAS.488.1941M} {488, 1941}

\bibitem[\protect\citeauthoryear{{McCracken} et~al.,}{{McCracken}
  et~al.}{2012}]{McCracken12}
{McCracken} H.~J.,  et~al., 2012, \mn@doi [\aap] {10.1051/0004-6361/201219507},
  \href {https://ui.adsabs.harvard.edu/abs/2012A&A...544A.156M} {544, A156}

\bibitem[\protect\citeauthoryear{{Mo}, {Jing}  \& {White}}{{Mo}
  et~al.}{1997}]{Mo97}
{Mo} H.~J.,  {Jing} Y.~P.,   {White} S.~D.~M.,  1997, \mn@doi [\mnras]
  {10.1093/mnras/284.1.189}, \href
  {https://ui.adsabs.harvard.edu/abs/1997MNRAS.284..189M} {284, 189}

\bibitem[\protect\citeauthoryear{{Mo}, {Mao}  \& {White}}{{Mo}
  et~al.}{1998}]{Mo98}
{Mo} H.~J.,  {Mao} S.,   {White} S. D.~M.,  1998, \mn@doi [\mnras]
  {10.1046/j.1365-8711.1998.01227.x}, \href
  {https://ui.adsabs.harvard.edu/abs/1998MNRAS.295..319M} {295, 319}

\bibitem[\protect\citeauthoryear{{Mosleh}, {Tacchella}, {Renzini}, {Carollo},
  {Molaeinezhad}, {Onodera}, {Khosroshahi}  \& {Lilly}}{{Mosleh}
  et~al.}{2017}]{Mosleh17}
{Mosleh} M.,  {Tacchella} S.,  {Renzini} A.,  {Carollo} C.~M.,  {Molaeinezhad}
  A.,  {Onodera} M.,  {Khosroshahi} H.~G.,   {Lilly} S.,  2017, \mn@doi [\apj]
  {10.3847/1538-4357/aa5f14}, \href
  {https://ui.adsabs.harvard.edu/abs/2017ApJ...837....2M} {837, 2}

\bibitem[\protect\citeauthoryear{{Mowla} et~al.,}{{Mowla}
  et~al.}{2019}]{Mowla19}
{Mowla} L.~A.,  et~al., 2019, \mn@doi [\apj] {10.3847/1538-4357/ab290a}, \href
  {https://ui.adsabs.harvard.edu/abs/2019ApJ...880...57M} {880, 57}

\bibitem[\protect\citeauthoryear{{Naab}, {Johansson}  \& {Ostriker}}{{Naab}
  et~al.}{2009}]{Naab09}
{Naab} T.,  {Johansson} P.~H.,   {Ostriker} J.~P.,  2009, \mn@doi [\apjl]
  {10.1088/0004-637X/699/2/L178}, \href
  {https://ui.adsabs.harvard.edu/abs/2009ApJ...699L.178N} {699, L178}

\bibitem[\protect\citeauthoryear{{Navarro} \& {Steinmetz}}{{Navarro} \&
  {Steinmetz}}{2000}]{Navarro00}
{Navarro} J.~F.,  {Steinmetz} M.,  2000, \mn@doi [\apj] {10.1086/309175}, \href
  {https://ui.adsabs.harvard.edu/abs/2000ApJ...538..477N} {538, 477}

\bibitem[\protect\citeauthoryear{{Obreschkow}, {Murray}, {Robotham}  \&
  {Westmeier}}{{Obreschkow} et~al.}{2018}]{Obreschkow18}
{Obreschkow} D.,  {Murray} S.~G.,  {Robotham} A.~S.~G.,   {Westmeier} T.,
  2018, \mn@doi [\mnras] {10.1093/mnras/stx3155}, \href
  {https://ui.adsabs.harvard.edu/abs/2018MNRAS.474.5500O} {474, 5500}

\bibitem[\protect\citeauthoryear{{Paulino-Afonso}, {Sobral}, {Buitrago}  \&
  {Afonso}}{{Paulino-Afonso} et~al.}{2017}]{PaulinoAfonso17}
{Paulino-Afonso} A.,  {Sobral} D.,  {Buitrago} F.,   {Afonso} J.,  2017,
  \mn@doi [\mnras] {10.1093/mnras/stw2933}, \href
  {https://ui.adsabs.harvard.edu/abs/2017MNRAS.465.2717P} {465, 2717}

\bibitem[\protect\citeauthoryear{{Peng}, {Ho}, {Impey}  \& {Rix}}{{Peng}
  et~al.}{2002}]{Peng02}
{Peng} C.~Y.,  {Ho} L.~C.,  {Impey} C.~D.,   {Rix} H.-W.,  2002, \mn@doi [\aj]
  {10.1086/340952}, \href
  {https://ui.adsabs.harvard.edu/abs/2002AJ....124..266P} {124, 266}

\bibitem[\protect\citeauthoryear{{Peng} et~al.,}{{Peng} et~al.}{2010}]{Peng10}
{Peng} Y.-j.,  et~al., 2010, \mn@doi [\apj] {10.1088/0004-637X/721/1/193},
  \href {https://ui.adsabs.harvard.edu/abs/2010ApJ...721..193P} {721, 193}

\bibitem[\protect\citeauthoryear{{R Core Team}}{{R Core
  Team}}{2020}]{R-Core-Team}
{R Core Team} 2020, R: A Language and Environment for Statistical Computing.
R Foundation for Statistical Computing, Vienna, Austria, \url
  {https://www.R-project.org/}

\bibitem[\protect\citeauthoryear{{Ravindranath} et~al.,}{{Ravindranath}
  et~al.}{2004}]{Ravindranath04}
{Ravindranath} S.,  et~al., 2004, \mn@doi [\apjl] {10.1086/382952}, \href
  {https://ui.adsabs.harvard.edu/abs/2004ApJ...604L...9R} {604, L9}

\bibitem[\protect\citeauthoryear{{Robotham}}{{Robotham}}{2016a}]{Robotham16a}
{Robotham} A. S.~G.,  2016a, {Celestial: Common astronomical conversion
  routines and functions} (\mn@eprint {ascl} {1602.011})

\bibitem[\protect\citeauthoryear{{Robotham}}{{Robotham}}{2016b}]{Robotham16b}
{Robotham} A. S.~G.,  2016b, {magicaxis: Pretty scientific plotting with
  minor-tick and log minor-tick support} (\mn@eprint {ascl} {1604.004})

\bibitem[\protect\citeauthoryear{{Robotham} \& {Obreschkow}}{{Robotham} \&
  {Obreschkow}}{2015}]{Robotham15}
{Robotham} A.~S.~G.,  {Obreschkow} D.,  2015, \mn@doi [\pasa]
  {10.1017/pasa.2015.33}, \href
  {https://ui.adsabs.harvard.edu/abs/2015PASA...32...33R} {32, e033}

\bibitem[\protect\citeauthoryear{{Robotham} et~al.,}{{Robotham}
  et~al.}{2014}]{Robotham14}
{Robotham} A.~S.~G.,  et~al., 2014, \mn@doi [\mnras] {10.1093/mnras/stu1604},
  \href {https://ui.adsabs.harvard.edu/abs/2014MNRAS.444.3986R} {444, 3986}

\bibitem[\protect\citeauthoryear{{Robotham}, {Taranu}, {Tobar}, {Moffett}  \&
  {Driver}}{{Robotham} et~al.}{2017}]{Robotham17}
{Robotham} A.~S.~G.,  {Taranu} D.~S.,  {Tobar} R.,  {Moffett} A.,   {Driver}
  S.~P.,  2017, \mn@doi [\mnras] {10.1093/mnras/stw3039}, \href
  {https://ui.adsabs.harvard.edu/abs/2017MNRAS.466.1513R} {466, 1513}

\bibitem[\protect\citeauthoryear{{Robotham}, {Davies}, {Driver}, {Koushan},
  {Taranu}, {Casura}  \& {Liske}}{{Robotham} et~al.}{2018}]{Robotham18}
{Robotham} A.~S.~G.,  {Davies} L.~J.~M.,  {Driver} S.~P.,  {Koushan} S.,
  {Taranu} D.~S.,  {Casura} S.,   {Liske} J.,  2018, \mn@doi [\mnras]
  {10.1093/mnras/sty440}, \href
  {https://ui.adsabs.harvard.edu/abs/2018MNRAS.476.3137R} {476, 3137}

\bibitem[\protect\citeauthoryear{{Scoville} et~al.,}{{Scoville}
  et~al.}{2007}]{Scoville07}
{Scoville} N.,  et~al., 2007, \mn@doi [\apjs] {10.1086/516751}, \href
  {https://ui.adsabs.harvard.edu/abs/2007ApJS..172..150S} {172, 150}

\bibitem[\protect\citeauthoryear{{Shen}, {Mo}, {White}, {Blanton}, {Kauffmann},
  {Voges}, {Brinkmann}  \& {Csabai}}{{Shen} et~al.}{2003}]{Shen03}
{Shen} S.,  {Mo} H.~J.,  {White} S. D.~M.,  {Blanton} M.~R.,  {Kauffmann} G.,
  {Voges} W.,  {Brinkmann} J.,   {Csabai} I.,  2003, \mn@doi [\mnras]
  {10.1046/j.1365-8711.2003.06740.x}, \href
  {https://ui.adsabs.harvard.edu/abs/2003MNRAS.343..978S} {343, 978}

\bibitem[\protect\citeauthoryear{{Trujillo} et~al.,}{{Trujillo}
  et~al.}{2006}]{Trujillo06}
{Trujillo} I.,  et~al., 2006, \mn@doi [\apj] {10.1086/506464}, \href
  {https://ui.adsabs.harvard.edu/abs/2006ApJ...650...18T} {650, 18}

\bibitem[\protect\citeauthoryear{{Trujillo}, {Conselice}, {Bundy}, {Cooper},
  {Eisenhardt}  \& {Ellis}}{{Trujillo} et~al.}{2007}]{Trujillo07}
{Trujillo} I.,  {Conselice} C.~J.,  {Bundy} K.,  {Cooper} M.~C.,  {Eisenhardt}
  P.,   {Ellis} R.~S.,  2007, \mn@doi [\mnras]
  {10.1111/j.1365-2966.2007.12388.x}, \href
  {https://ui.adsabs.harvard.edu/abs/2007MNRAS.382..109T} {382, 109}

\bibitem[\protect\citeauthoryear{{Trujillo}, {Ferreras}  \& {de La
  Rosa}}{{Trujillo} et~al.}{2011}]{Trujillo11}
{Trujillo} I.,  {Ferreras} I.,   {de La Rosa} I.~G.,  2011, \mn@doi [\mnras]
  {10.1111/j.1365-2966.2011.19017.x}, \href
  {https://ui.adsabs.harvard.edu/abs/2011MNRAS.415.3903T} {415, 3903}

\bibitem[\protect\citeauthoryear{{White} \& {Frenk}}{{White} \&
  {Frenk}}{1991}]{White91}
{White} S. D.~M.,  {Frenk} C.~S.,  1991, \mn@doi [\apj] {10.1086/170483}, \href
  {https://ui.adsabs.harvard.edu/abs/1991ApJ...379...52W} {379, 52}

\bibitem[\protect\citeauthoryear{{Xie}, {Guo}, {Cooper}, {Frenk}, {Li}  \&
  {Gao}}{{Xie} et~al.}{2015}]{Xie15}
{Xie} L.,  {Guo} Q.,  {Cooper} A.~P.,  {Frenk} C.~S.,  {Li} R.,   {Gao} L.,
  2015, \mn@doi [\mnras] {10.1093/mnras/stu2487}, \href
  {https://ui.adsabs.harvard.edu/abs/2015MNRAS.447..636X} {447, 636}

\bibitem[\protect\citeauthoryear{{York} et~al.,}{{York} et~al.}{2000}]{York00}
{York} D.~G.,  et~al., 2000, \mn@doi [\aj] {10.1086/301513}, \href
  {https://ui.adsabs.harvard.edu/abs/2000AJ....120.1579Y} {120, 1579}

\bibitem[\protect\citeauthoryear{{van der Wel}, {Holden}, {Zirm}, {Franx},
  {Rettura}, {Illingworth}  \& {Ford}}{{van der Wel}
  et~al.}{2008}]{vanderWel08}
{van der Wel} A.,  {Holden} B.~P.,  {Zirm} A.~W.,  {Franx} M.,  {Rettura} A.,
  {Illingworth} G.~D.,   {Ford} H.~C.,  2008, \mn@doi [\apj] {10.1086/592267},
  \href {https://ui.adsabs.harvard.edu/abs/2008ApJ...688...48V} {688, 48}

\bibitem[\protect\citeauthoryear{{van der Wel} et~al.,}{{van der Wel}
  et~al.}{2012}]{vanderWel12}
{van der Wel} A.,  et~al., 2012, \mn@doi [\apjs] {10.1088/0067-0049/203/2/24},
  \href {https://ui.adsabs.harvard.edu/abs/2012ApJS..203...24V} {203, 24}

\bibitem[\protect\citeauthoryear{{van der Wel} et~al.,}{{van der Wel}
  et~al.}{2014}]{vanderWel14}
{van der Wel} A.,  et~al., 2014, \mn@doi [\apj] {10.1088/0004-637X/788/1/28},
  \href {https://ui.adsabs.harvard.edu/abs/2014ApJ...788...28V} {788, 28}

\makeatother
\end{thebibliography}

\end{document}